\documentstyle[graphicx]{elsart}

\newcommand{\beq}{\begin{equation}}
\newcommand{\eeq}{\end{equation}}
\newcommand{\bqa}{\begin{eqnarray}}
\newcommand{\eqa}{\end{eqnarray}}

\newcommand{\sump}{\mathop{{\sum}'}}

\begin{document}
\begin{frontmatter}
\title{Electromagnetic transitions between
giant resonances within a continuum-RPA approach}

\author[TUEB,KVI,MEPI]{ V.A.~Rodin \thanksref{e-mail}}
and
\author[KVI]{A.E.L.~Dieperink}
\address[TUEB]{Institut f\"ur Theoretische Physik der Universit\"at T\"ubingen,
Auf der Morgenstelle 14, D-72076 T\"ubingen, Germany}
\address[KVI]{Kernfysisch Versneller Institute, NL-9747AA Groningen,
The Netherlands}
\address[MEPI]{Department of Theoretical Nuclear Physics,
Moscow State Engineering and Physics Institute,115409 Moscow, Russia}
\thanks[e-mail]{E-mail: vadim.rodin@uni-tuebingen.de}

\begin{abstract}
A general continuum-RPA approach is developed to describe
electromagnetic transitions between giant resonances.  Using a diagrammatic
representation for the three-point Green's function, an expression for
the transition amplitude  is derived which
allows one to incorporate  effects of
mixing of single and double giant resonances as well as to take the entire
basis of particle-hole states into consideration.
The radiative widths for E1 transition between
the charge-exchange spin-dipole giant resonance and Gamow-Teller states
are calculated for $^{90}$Nb and $^{208}$Bi.
The importance of the mixing is stressed.
\begin{flushleft}
\end{flushleft}
\begin{flushleft}
\end{flushleft}

\end{abstract}
\end{frontmatter}
\section{Introduction}
Since the very beginning of nuclear physics, electromagnetic
transitions between nuclear states have been known to be a
sensitive tool to investigate structure of the nuclear wave
functions. In recent years significant theoretical efforts have
been devoted to describe intensities of the radiative transitions
between charge-exchange giant resonances (GR) in nuclei
{\cite{VanG95}-\cite{Rod00}}. The interest in the problem is
related to the possibility to obtain  corresponding
experimental data from the ($^3$He,$t\gamma$)-reaction cross
sections. A confrontation of the data with the calculation results would be
especially interesting in view of the theoretical speculations on
the possible enhancement of the transition intensities
{\cite{VanG95}-\cite{Sag00}}.

It should be mentioned that conclusions on the enhancement have
been based upon the consideration of the relevant sum rules.
However, the main contribution to the sum rules seems to come from
the nuclear states of 2 particle - 2 hole type (double GR). For
instance, the dipole sum rule for the Gamow-Teller resonance (GTR)
should be mainly exhausted by the giant dipole resonance (GDR)
built on the top of the GTR, i.e. a double GR, but not by the
spin-dipole resonance (SDR), a single GR. In this connection the
problem of a correct description for the mixing of single and
double GR arises, because it might change significantly the
transition intensities. The relevance of this problem is indicated by
the observation that the calculations of the radiative transition
intensities within TDA \cite{VanG95}-\cite{Sag00} and RPA
\cite{Rod00} when taking only particle-hole structure of GRs into
consideration have strongly underestimated the corresponding sum
rules. It is noteworthy,
that in recent works \cite{Pon98} the importance of taking into account
the GDR admixture has been demonstrated to describe correctly
the decay amplitude of the first $1^-$ (two-phonon type) state to the ground state.

In the present work a general expression for the amplitude of the
radiative transitions between different GRs is obtained for
closed-shell nuclei within the continuum-RPA using a diagrammatic
representation for the three-point Green's function. The
expression allows us to take the entire basis of particle-hole
states into consideration as well as to incorporate the important effects of
mixing of single and double GR.

We apply the general approach developed in this paper to calculate
the intensity of E1 transition between the SDR and GT states. The
calculation results for $^{90}$Nb and $^{208}$Bi nuclei differ
markedly from the previous ones {\cite{VanG98}-\cite{Rod00}},
mainly due to the effects of mixing of the single and double GR.
The corresponding experimental data for the latter nucleus is to appear
soon~\cite{Kr02}.

\section{The amplitude of electromagnetic transitions between giant resonances}

We start with the well-known formula for the partial width
corresponding to a radiative transition between initial
$|1\rangle$ and final $|2\rangle$ nuclear states with the excitation
energies $E_1$ and $E_2$, respectively,:
\beq
\Gamma_\gamma^{(1\to 2)}=C_\gamma \sum_{m_2\mu} \left|
\langle 2,m_2|\hat {\cal Q}_{k\mu} |1,m_1\rangle\right|^2= C_\gamma
\frac{\left|\langle 2\|\hat {\cal Q}_{k} \|1\rangle\right|^2}{2J_2+1}.
\label{eq1}
\eeq
Here, $\hat {\cal Q}_{k\mu}=\sum_{a} Q_{k\mu}(x_a)$ is a
single-particle multipole operator, $\langle 2\|\hat {\cal Q}_{k}
\|1\rangle$ is the reduced matrix element, $x$ is the nucleon coordinate
(including spin and isospin variables), $m$ is the
angular momentum projection, $C_\gamma$ is an appropriate
dimensional coefficient. In the case of E1 transitions, 
considered in the present paper to apply the general approach (see Sect.~3), 
$Q_{1\mu}= D_{\mu}=-\frac12 e\, r\, Y_{1\mu}(\vec
n)\,\tau^{(3)}$ is the isovector electric dipole operator and
$C_\gamma=\frac{16\pi}{9}\left(\frac{E_\gamma}{\hbar c}\right)^3$
with $E_\gamma=E_1-E_2$ being the gamma-quantum energy.

Let $|1\rangle$ and $|2\rangle$ be states of the particle-hole (p-h) type
so that
their structure can be described within the RPA. Expression for
the transition amplitude $\langle 2\|\hat {\cal Q}_{k}
\|1\rangle$ can be obtained in Green function technique used in
the theory of finite Fermi-systems~\cite{Mig83} to describe the
Fermi-system response to a single-particle probing operator. Let
us consider the diagrams for the 3-point Green function
(3-point vertex function) describing the transition amplitude of
the system under the action of the external field $\hat {\cal Q}$ (Fig.~1).

\begin{figure}[h]
 \begin{center}
 \includegraphics[width=8.6cm,clip]{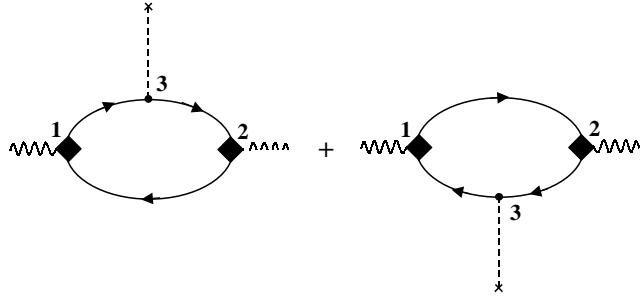}
\caption{ Three-point Green's function describing the amplitude of the 
transition between the RPA states 1 and 2.
The solid line presents the nucleon,
the dashed one
the single-particle external field $Q_k(x)$, and the wavy lines
the transition densities.}
 \end{center}
\end{figure}

The p-h structure of the state 1~(2) enters these graphs by means of
the transition fields $v_{1(2)}(x)=\int F(xx')\rho_{1(2)}(x')\,dx'$ 
with $\rho_{1(2)}(x)$ and $F(xx')$ being the transition densities
and the particle-hole interaction (depicted by the full diamonds), 
respectively.
The transition fields are proportional to the residues in the poles
of the effective operator
:
$\tilde
V_{1(2)}(\omega)\simeq
\displaystyle\frac{M_{1(2)}v_{1(2)}}{\omega-E_{1(2)}}$, with
$\tilde V$ satisfying the equation{~\cite{Mig83}}:
\beq
\tilde V(x,\omega)=V(x)+\int F(xx')A(x'x'',\omega)
\tilde V(x'',\omega)\,dx'dx'',
\label{effiel}
\eeq
Here, $A(xx',\omega)$ is the free particle-hole propagator,
carrying the quantum numbers of the appropriate p-h state,
the probing single-particle operator $V_{1(2)}(x)$ has the matrix
element $M_{1(2)}=\langle 1(2)|\hat {\cal V}_{1(2)}|0\rangle$
($\hat {\cal V}=\sum_{a} V(x_a)$).

The matrix element corresponding  to the graphs is:
\bqa
&& \label{amplit1}
\hskip-1.cm \langle 2\| \hat {\cal Q}_k \| 1 \rangle =
(-)^{J_1-k}\sum\limits_{345}
\left\{\!\arraycolsep=0.05em\begin{array}{lll}J_2& k & J_1\\ j_3& j_4& j_5
\end{array}\!\!\right\}
\left[(-)^{J_1+J_2+k}\langle 4\|v_2^+\|5\rangle\langle 5\|Q_k\|3\rangle\langle 3\|v_1\|4\rangle
\phantom{J^{J}}\right.\\
&& \hskip-1.cm \times\left(\frac{n_4 (1-n_5)(1-n_3)}
{(\varepsilon_4+E_{2}-\varepsilon_5)(\varepsilon_4+E_{1}-\varepsilon_3)}
- \frac{(1-n_4)n_5n_3}
{(\varepsilon_4+E_{2}-\varepsilon_5)(\varepsilon_4+E_{1}-\varepsilon_3)}
\right)\nonumber\\
&&\hskip-1.cm -\langle 4\|v_1\|3\rangle\langle 3\|Q_k\|5\rangle \langle 5\|v_2^+\|4\rangle
\left.\left(\frac{(1-n_4)n_5n_3}
{(\varepsilon_4-E_{2}-\varepsilon_5)(\varepsilon_4-E_{1}-\varepsilon_3)}
-\frac{n_4(1-n_5)(1-n_3)}
{(\varepsilon_4-E_{2}-\varepsilon_5)(\varepsilon_4-E_{1}-\varepsilon_3)}
\right)\right],\nonumber
\eqa
with $n_\lambda$ and $\varepsilon_\lambda$ being occupation numbers and energies of the
single-particle states, respectively ($\lambda$ is the set of the single-particle
quantum numbers, $(\lambda)=\{lj\}$). The definition
of the reduced single-particle matrix elements $\langle \lambda\| Q_k \|\mu
\rangle$ by the Wigner-Eckart theorem is taken in accordance with
{Ref.~\cite{bm}}.

The Eq.~(\ref{amplit1})
can be transformed to express the transition amplitude
in terms of the RPA $X$ and $Y$ amplitudes (see, e.g., Refs.~\cite{Pon98,Sol}):
\bqa
\label{amplXY}
& \langle 2\|\hat {\cal
Q}_{k}\|1\rangle= \sqrt{(2J_1+1)(2J_2+1)}\sum\limits_{345}
(-)^{J_1+1+j_4+j_3}
\left\{\!\arraycolsep=0.05em\begin{array}{lll} J_2& k& J_1\\ j_3& j_4& j_5
\end{array}\!\!\right\}\nonumber\\
& \times\left[(-)^{J_1+J_2+k}\langle 5\|Q_{k} \|3\rangle
\left( X^{(1)}_{34}X^{(2)}_{54} +
(-)^{J_1+J_2+j_3+j_5}Y^{(1)}_{43}Y^{(2)}_{45} \right)\right.\nonumber\\
& \left.-\langle 3\|Q_{k} \|5\rangle
\left(X^{(1)}_{43}X^{(2)}_{45} + (-)^{J_1+J_2+j_3+j_5}Y^{(1)}_{34}Y^{(2)}_{54}
\right)\right]
\eqa

with the use of the following representations for the $X$ and $Y$:
\bqa
& X^{(1)}_{21}=(-)^{j_2-j_1+J_1+1}\frac{\langle 2\|v_1\|1\rangle}
{\sqrt{2J_1+1}}\frac{n_1(1-n_2)}{(\varepsilon_1+E_{1}-\varepsilon_2)}\\
& Y^{(1)}_{21}=(-)^{j_2-j_1+1}\frac{\langle 1\|v_1\|2\rangle}
{\sqrt{2J_1+1}}\frac{n_1(1-n_2)}{(\varepsilon_1-E_{1}-\varepsilon_2)}
\label{XY}
\eqa

One sees in Fig.~1 that there is an asymmetry between the vertex 1,2 ("dressed") on the  one hand and
vertex 3 ("undressed") on the other. This is because we have not
incorporated effects of particle-hole interaction in channel 3, or, in other
words, we neglected the effect of the virtual excitation of the corresponding
giant multipole  resonance. This effect has been well-known to play
a crucial role in describing the neutron radiative capture
since the model of direct-semidirect capture was suggested~\cite{brown} (see
also Ref.~\cite{Rod00a} for more recent developments).
This additional contribution which takes into account virtual excitation of
the  giant resonance  
is shown in Fig.~2.

\begin{figure}[h]
 \begin{center}
 \includegraphics[width=8.6cm,clip]{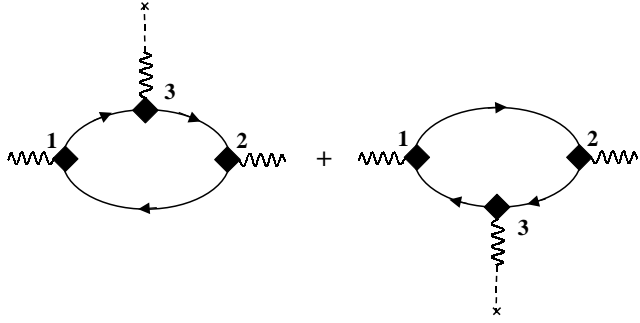}
 \caption{The generalized three-point Green's function
which takes into account the virtual excitation 
 of multipole giant resonance in the channel 3.}
 \end{center}
\end{figure}
The wavy line in the channel 3 depictures the total p-h propagator
$\tilde A$ describing off-shell excitation of all p-h states (including
the corresponding GR) with appropriate spin and parity.
Using the fact that $\tilde A Q=A\tilde Q$, one can see that the sum
of the graphs of Figs.1-2 is equivalent to the graphs of Fig.1 only
if one replaces the external field $Q_k(x)$ by the corresponding
effective field $\tilde Q_k(x,\omega)$ satisfying Eq.(\ref{effiel}).

The Eq.(\ref{amplit1}) with the substitution $Q_k(x)\to\tilde Q_k(x,\omega)$
can be transformed further with the use of the spectral decomposition for the
Green's function of the single-particle radial Schr\"odinger equation
$g_{(\lambda)}(rr',\varepsilon)=
\displaystyle\sump\limits_{\lambda}\frac{\chi_\lambda(r)\chi_\lambda(r')} {\varepsilon-\varepsilon_\lambda}$
to include explicitly the single-particle continuum:
\bqa
\label{amplit2}
&\langle 2\| \hat {\cal Q}_k\| 1 \rangle =(-)^{J_1-k}
\left[ \sum\limits_{4}n_4\left((-)^{J_1+J_2+k}G_4(E_{1},E_{2})+G_4^T(-E_{1},-E_{2})\right)
\right.\nonumber\\
&\left. -\sum\limits_{3,5}n_3n_5\left((-)^{J_1+J_2+k}G_{3,5}(E_{1},E_{2})+G_{3,5}^T(-E_{1},-E_{2})\right)
\right], \\
& \nonumber\\
& \nonumber\\
& G_4(E_{1},E_{2})=\displaystyle
\sum\limits_{(3)(5)} t_{(3)(4)(5)}(4|v_{2}^+ \left( g_{(5)}(\varepsilon_4+E_{2})
- \sump\limits_{5} n_5 \frac{|5)(5|}{\varepsilon_4+E_{2}-\varepsilon_5} \right)
\nonumber\\
&\times\tilde Q_k
\displaystyle \left( g_{(3)}(\varepsilon_4+E_{1}) - \sump\limits_{3} n_3 \frac{|3)(3|}
{\varepsilon_4+E_{1}-\varepsilon_3} \right)v_{1}|4),
\nonumber\\
& G_{3,5}(E_{1},E_{2})= \sum\limits_{(4)} t_{(3)(4)(5)}
\displaystyle\frac{(5|\tilde Q_k|3)}{\varepsilon_5-\varepsilon_3+E_{1}
-E_{2}}(3|v_{1} (g_{(4)}(\varepsilon_3-E_{1})- g_{(4)}(\varepsilon_5-E_{2}))v_{2}^+|5),
\nonumber\\
& t_{(3)(4)(5)}=\left\{\!\arraycolsep=0.05em\begin{array}{lll} J_2& k& J_1\\ j_3& j_4& j_5
\end{array}\!\!\right\}
\langle (4)\|v_2^+\|(5)\rangle\langle (5)\|Q_k\|(3)\rangle\langle (3)\|v_1\|(4)\rangle,
\nonumber
\eqa
where we have decomposed the reduced matrix elements
$\langle \lambda\|v\|\mu\rangle \equiv \langle (\lambda)\|v\|(\mu)\rangle (\lambda|v|\mu)$ into
its spin-angular $\langle (\lambda)\|v\|(\mu)\rangle$ and radial
$(\lambda|v|\mu)=\int \chi_\lambda(r)v(r)\chi_\mu(r)\,dr$ parts. The
radial integrals (for brevity omitted) can be easily recovered, for instance,
\bqa
&(4|v_{2}^+ g_{(5)}(\varepsilon_4+E_{2}) \tilde Q_k
g_{(3)}(\varepsilon_4+E_{1})v_{1}|4)= \nonumber\\
&\int \chi_4(r)v_2^+(r)
g_{(5)}(rr',\varepsilon_4+E_{2})\tilde Q_k(r')
g_{(3)}(r'r'',\varepsilon_4+E_{1})v_{1}(r'')\chi_4(r'')\,drdr'dr''.
\nonumber
\eqa
The superscript $T$ means transposition and $\sump$ runs over
single-particle states with the same $l$ and $j$.

The Eq.~(\ref{amplit2}) gives the general expression for the amplitude of the
electromagnetic transitions within the CRPA. Apart from the single-particle continuum, it
contains also contribution from the double phonon configurations. To see this, let us decompose
the amplitude into "direct" and "semi-direct" parts according to the partition
$\tilde Q_k=Q_k+(\tilde Q_k - Q_k)$.
The first, direct, part describes the amplitude in the single-phonon space and
it has been only taken into account in all previous theoretical considerations{~\cite{VanG95}-\cite{Rod00}}.
As for the semi-direct part, the following approximate representation
holds:
\beq
\tilde Q_k(x,\omega) - Q_k(x) \simeq \sum_i\frac{M_{i}q_i(x)}{\omega-\omega_i}.
\label{1.15}
\eeq
with $q_i(x)$ being the transition fields for the doorway states forming the GDR.
Thus, one can see, that the semi-direct part
originates from admixtures of double-phonon configurations, like
$|2\rangle +\sum_i\alpha_i|1\rangle\otimes|i\rangle$.
The contribution to the transition amplitude is determined by both the matrix elements
$M_{i}=\langle i|\hat {\cal Q}_k|0\rangle$ (states $|1\rangle$ or $|2\rangle$ serve
as spectators in these electromagnetic transitions) and the
admixture amplitudes $\alpha_i$.

It can be seen from Eqs.~(\ref{amplit2},\ref{1.15}) that the mixing is described within
first order perturbation theory with respect to the particle-hole
interaction. Nevertheless, the approximation is satisfied in most cases
due to the small values of the mixing amplitudes, that can be also seen from the calculated
transition amplitudes being much less than the amplitude between the GDR and the ground state.

\section{E1 transitions between the spin-dipole and Gamow-Teller resonances.}
As an application of the approach described in the previous section, we
calculate  intensities
and branching ratios for E1 decay of the charge-exchange
spin-dipole resonance to some Gamow-Teller states (GTR and low-lying satellites)
identified in Ref.~\cite{Kr01}.
To calculate the GT and SD strength distribution, we
use the CRPA approach of Refs.{~\cite{Rod00,Moukh99,Rod01}}.
The phenomenological isoscalar part of the
nuclear mean field and the zero-range Landau-Migdal particle-hole interaction are the ingredients
of the approach along with some selfconsistency conditions.
The approach has allowed one to reproduce very well the
experimental GTR energies in $^{90}$Nb and $^{208}$Bi nuclei{~\cite{Rod01}}.
For each spin $J^\pi_{SDR}=(0^-,1^-)$ of the SD components
we choose just one state with the maximal strength to be the SDR, lying at (26.6, 24.1) MeV
in $^{208}$Bi and at (22.4, 20.1) MeV in $^{90}$Nb, and for $J^\pi_{SDR}=2^-$
we choose three states with the maximal excitation energies, lying at (20.0, 21.3, 23.1) MeV
in $^{208}$Bi and at (13.5, 14.4, 17.4) MeV in $^{90}$Nb.

The details of the calculation of the effective fields for the SD and GT cases can be found in
Refs.{~\cite{Rod00,Moukh99,Rod01}}, and for the isovector dipole one - in Ref.{~\cite{Rod00a}}. In the
latter case we take the velocity dependent particle-hole forces along with
the isovector Landau-Migdal interaction that allows us
to reproduce in the calculations both the experimental GDR energy and
the observed excess of the energy-weighted sum rule over the TRK one.
Because some transition energies
are rather close to the GDR energy, the influence of the GDR spreading becomes important.
We use the substitution $\omega \to \omega + {\rm i}\Gamma(\omega)/2$
in Eq.~(\ref{effiel}) to simulate the effect of GDR spreading
(the mean energy-dependent doorway spreading width $\Gamma(\omega)$ has been
chosen as in Ref.~\cite{Rod00a}).

We calculate the reduced matrix elements $\langle GTR\|\hat {\cal D}_{k} \|SDR,J\rangle$
according to Eq.~(\ref{amplit2}) and Eq.~(\ref{amplit1}) (with and
without taking into account the virtual GDR excitation, respectively),
and then the corresponding radiative widths $\Gamma_\gamma$ and $\Gamma_\gamma^0$
according to Eq.~(\ref{eq1}). The calculated widths for the E1 decay of the above-mentioned
SD states to the GT states at the excitation energy $E_{GT}$ are listed in  Table 1.
The corresponding branching ratios are calculated as the ratio $\Gamma_\gamma/\Gamma^\downarrow$
with $\Gamma^\downarrow$ being the mean doorway-state spreading width
chosen to reproduce the experimental total SDR width in the strength-function calculations
($\Gamma^\downarrow=4.7$ MeV for $^{208}$Bi \cite{Moukh99}, $\Gamma^\downarrow=4.$ MeV for
$^{90}$Nb).

\begin{table}[htb]
\caption{Calculated radiative widths and branching ratios for the E1
transitions between SDR and GT states in $^{90}$Nb and $^{208}$Bi
nuclei without 
($\Gamma_\gamma^0,b_\gamma^0$) and with ($\Gamma_\gamma,b_\gamma$)
taking into account the virtual GDR excitation, respectively.}
\label{tab1}
{\begin{center}
\begin{tabular}{|c|c|c|c|c|c|c|c|}
\hline
 & & \multicolumn{3}{|c|}{}& \multicolumn{3}{|c|}{}\\
 & {$E_{GT}$, MeV} &
\multicolumn{3}{|c|}{$\Gamma_\gamma^0$, keV} & \multicolumn{3}{|c|}{$\Gamma_\gamma$, keV} \\
Nucleus & & \multicolumn{3}{|c|}{($b_\gamma^0$, $\times10^{-4}$)} & \multicolumn{3}{|c|}{($b_\gamma$, $\times10^{-4}$)} \\
\cline{3-8}
&\multicolumn{1}{|r|}{$ \left/ J^\pi_{SDR} \right.$}
& 0$^-$ & 1$^-$ & 2$^-$ & 0$^-$ & 1$^-$ & 2$^-$ \\
\hline
 & 8.9
 & 3.40 & 1.03 & 0.10 & 3.68 & 0.02 & 0.008 \\
$^{90}$Nb &  & (8.5) & (2.6) & (0.3) & (9.2) & (0.05) & (0.02) \\
\cline{2-8}
 & 0.0
 & 0.21 & 0.10 & 0.08 & 1.62 & 0.13 & 0.37 \\
  &  & (0.5) & (0.2) & (0.2) & (4.0) & (0.3)& (0.9) \\
\hline
& 15.5
& 2.28 & 0.70 & 0.09 & 6.49 & 0.03 & 0.008 \\
 &  & (4.9) & (1.5) & (0.2) & (13.8) & (0.05) & (0.02) \\
\cline{2-8}
 & 9.3
 & 0.15 & 0.03 & 0.14 & 0.80 & 0.43 & 0.22 \\
 &  & (0.3) & (0.05) & (0.3) & (1.7) & (0.9) & (0.5) \\
\cline{2-8}
 & 8.1
 & 0.24 & 0.02 & 0.04 & 1.05 & 0.26 & 0.16 \\
 &  & (0.5) & (0.03) & (0.08) & (2.3) & (0.5) & (0.3) \\
\cline{2-8}
$^{208}$Bi & 6.4
& 0.23 & 0.007 & 0.001 & 0.58 & 0.04 & 0.03 \\
 &  & (0.5) & (0.01) & (0.002) & (1.2) & (0.1) & (0.06) \\
\cline{2-8}
 & 4.9
 & 1.05 & 0.0002 & 0.002 & 2.36 & 0.003 & 0.01 \\
 &  & (2.2) & (0.0005) & (0.003) & (5.0) & (0.007) & (0.03) \\
\cline{2-8}
 & 4.5
 & 0.50 & 0.0003 & 0.001 & 1.24 & 0.004 & 0.005 \\
 &  & (1.0) & (0.001) & (0.003) & (2.6) & (0.01) & (0.01) \\
\cline{2-8}
 & 3.8
 & 0.58 & 0.004 & 0.006 & 1.19 & 0.02 & 0.02 \\
 &  & (1.2) & (0.01) & (0.01) & (2.5) & (0.03) & (0.04) \\
\cline{2-8}
 & 1.2
 & 0.1 & 0.19 & 0.02 & 0.21 & 0.42 & 0.04 \\
 &  & (0.2) & (0.4) & (0.04) & (0.4) & (0.9) & (0.1) \\
\hline
\end{tabular}
\end{center}
}
\end{table}

As can be seen from the calculation results listed in the Table 1, taking
into account of the virtual GDR excitation affects
the results drastically. In the cases
where the transition energy is well below the GDR energy,
a destructive interference between the direct and semidirect parts
of the amplitude always occurs, in agreement with
both the qualitative description in terms of the effective charge
\cite{bm} and previous quantitative consideration of Ref.~\cite{Pon98}.
In the cases when the transition energy is close to the GDR energy or
exceeds it, the opposite situation takes place and one can see a significant
enhancement of the corresponding amplitudes. This circumstance along
with the steep $E_\gamma^3$ dependence of the E1 radiative width leads to
the fact that the widths for SDR decay to some low-lying GT states can
be comparable to those for SDR $\to$ GTR E1 transitions.
Some of the calculated branching ratios are of order of $10^{-3}$,
that seem to be accessible in experiments.
In fact, the statistical background hinders detection of SDR $\to$ GTR decay,
whereas it is small enough
to allow detection of SDR decay to low-energy GT states which has been already observed in
($^3$He,$t\gamma$) experiments on $^{208}$Pb~\cite{Kr02}.
For instance, strong gamma transitions have been observed to the unresolved GT states
with E=3.8, 4.5, 4.9 MeV~\cite{Kr02}. The energy of these gamma transitions (about 14 MeV) is very close to
the energy of the GDR, and the predicted branching ratio ($\sim 10^{-3}$) can explain
the observed transitions.
However, we have to mention that the present calculations overestimate the mean energy of the SDR
by about 2 MeV~\cite{Moukh99}. Bearing in mind strong energy-dependent renormalization of the
transition amplitude found in this paper, this can lead to
a noticeable difference between the calculated and observed branching ratios.
We have not also taken the finite widths of the GTR and the SDR into consideration.

Special interest for testing the model lies
in the E1 transition from the GTR to a 2.9 MeV 2$^-$ SD state
in $^{208}$Bi, which was observed recently at KVI~\cite{Kr02}.
The latter state has presumably a ($\pi 1h_{9/2})(\nu
1i_{13/2})^{-1}$  particle-hole structure and has a predicted excitation energy of 2.4 MeV.
For this E1 transition we have found a large branching ratio $b_\gamma=4.3\cdot 10^{-3}$
due to the fact that
the transition energy almost exactly coincides with the GDR energy,
therefore we obtain maximal enhancement.

Finally, it is also worth to mention the sensitivity of the radiative widths to the spin of the SDR
components due to their different energy position
that could be in principle used to get experimentally information on their energy
splitting.

\section{Conclusion}
A general continuum-RPA approach has been developed in the present paper
to describe the intensities
of electromagnetic transitions between different giant resonances.
We have used a diagrammatic representation for the three-point Green's function
to derive the expression for the transition amplitude. The
expression has allowed us to take the entire basis of particle-hole
states into consideration as well as to incorporate important effects of
mixing of single and double giant resonances.
The radiative widths for E1 transition between the SDR and GT states
(including GTR) have been calculated for $^{90}$Nb and $^{208}$Bi nuclei.
Some of the widths seem to be large enough to be accessible in experiments.

\section*{Acknowledgments}
Authors are grateful to Prof.~M.N.~Harakeh and Dr.~A.~Krasznahorkay for useful discussions and
suggestions as well as to the Referee for putting forward a proposal improving the figures' content.
This work was supported in part by the Nederlandse Organisatie voor
Wetenschapelijk Onderzoek.
V.A.R. would like also to thank the Graduiertenkolleg "Hadronen im Vakuum,
in Kernen und Sternen" (GRK683) for supporting his stay in T\"ubingen and
Prof.~A.~F\"a{\ss}ler for hospitality.

\end{document}